\documentclass[twocolumn,aps,prb,floatfix,showpacs]{revtex4}

\usepackage{graphicx}

\begin{document}

\title{Raising Bi-O bands above the Fermi energy level of hole-doped
Bi$_2$Sr$_2$CaCu$_2$O$_{8+\delta}$ and other cuprate superconductors}

\author{Hsin Lin, S. Sahrakorpi, R.S. Markiewicz and A. Bansil}

\affiliation{Physics Department, Northeastern University, Boston MA
02115, USA }

\begin{abstract}

   The Fermi surface (FS) of Bi$_2$Sr$_2$CaCu$_2$O$_{8+\delta}$
   (Bi2212) predicted by band theory displays Bi-related pockets
   around the $(\pi,0)$ point, which have never been observed
   experimentally. We show that when the effects of hole doping either
   by substituting Pb for Bi or by adding excess O in Bi2212 are
   included, the Bi-O bands are lifted above the Fermi energy ($E_F$)
   and the resulting first-principles FS is in remarkable accord with
   measurements.  With decreasing hole-doping the Bi-O bands drop
   below $E_F$ and the system self-dopes below a critical hole
   concentration. Computations on other Bi- as well as Tl- and
   Hg-based compounds indicate that lifting of the cation-derived band
   with hole doping is a general property of the electronic structures
   of the cuprates.

\end{abstract}

\date{\today}


\maketitle

First-principles band theory computations on the cuprates have become
a widely accepted tool for gaining insight into their electronic
structures, spectral properties, Fermi surfaces (FS's), and as a
starting point for constructing theoretical models for incorporating
strong correlation effects beyond the framework of the local-density
approximation (LDA) underlying such calculations
\cite{Pickett1989,Pavarini2001,Bansil1999,Markiewicz2005}. For
example, in the double layer Bi-compound
Bi$_2$Sr$_2$CaCu$_2$O$_{8+\delta}$ (Bi2212) $-$ perhaps the most
widely investigated cuprate $-$ the LDA generated band
structure\cite{Hybertsen1988,Bellini2004} is commonly invoked to
describe the doped metallic state of the system. Band theory however
clearly predicts the FS of Bi2212 to contain a FS pocket around the
antinodal point $M(\pi,0)$ as a Bi-O band drops below the Fermi energy
($E_F$), but such FS pockets have never been observed
experimentally\cite{damascelli03}. This `Bi-O pocket problem' is quite
pervasive and occurs in other Bi-compounds.\cite{Singh1995} Similarly,
Tl- and Hg-compounds display cation-derived FS pockets, presenting a
fundamental challenge for addressing on a first-principles basis
issues related to the doping dependencies of the electronic structures
of the cuprates.

In this Letter, we show how the cation-derived band responsible for
the aforemenentioned FS pockets is lifted above $E_F$ when hole doping
effects are properly included in the computations. Detailed results
for the case of Bi2212 are presented, where hole doping is generated
either by substituting Pb for Bi or by adding excess oxygen in the
Bi-O planes.  With 20\% Pb doping in the orthorhombic crystal
structure, the Bi-O band lies $\approx$ 1 eV above $E_F$ and the
remaining bonding and antibonding FS sheets are in remarkable accord
with the angle-resolved photoemission (ARPES) measurements on an
overdoped Bi2212 single crystal\cite{Bogdanov2001}. Below a critical
hole doping level, the Bi-O band falls below $E_F$ and, as a result of
this self-doping effect, further reduction in the hole doping level no
longer reduces the number of holes in the CuO$_2$ layers. We argue
that the underlying mechanism at play here is that hole doping reduces
the effective positive charge in the Bi-O donor layers, which then
reduces the tendency of the electrons to `flow back' and self-dope the
material.  We have also carried out computations on a number of
related compounds, including monolayer and trilayer Bi-compounds
[Bi$_2$Sr$_2$CuO$_{6+\delta}$ (Bi2201) and
Bi$_2$Sr$_2$Ca$_2$Cu$_3$O$_{10+\delta}$ (Bi2223)], and the Tl- and
Hg-based compounds, and we find that the lifting of the cation-derived
band with hole doping is a generic property of many families of
cuprates.

Concerning technical details, we have employed both the
Korringa-Kohn-Rostoker (KKR) and linearized augmented plane wave
(LAPW) band structure methodologies where we treat all electrons in
the system self-consistently and consider the full crystal potential
without the muffin-tin approximation.\cite{KKR1999,wien2k} The KKR
scheme is well-known to be particularly suited for a first-principles
treatment of the electronic structure of substitutionally disordered
alloys. Pb substitution on the Bi sites was considered within the
framework of the virtual crystal approximation (VCA), where the Bi
nuclear charge $Z$ is replaced by the average of the Bi and Pb charges
of what may be thought of as an `effective' Bi/Pb atom, but otherwise
the band structure problem is solved fully self-consistently
maintaining the charge neutrality of the system. The VCA is expected
to be a good approximation in this case since the effective disorder
parameter for the Bi-O states, given by $\Delta/W$, where $\Delta$ is
the splitting of the Bi-O and Pb-O bands in Bi2212 and Pb2212,
respectively, and $W$ is the band width, is estimated to be $\sim$
0.3, so that the system is far from being in the split-band
limit.\cite{VCA} We have also carried out superlattice computations by
substituting two Bi atoms by Pb in the orthorhombic Bi2212 as well as
KKR-CPA (coherent potential approximation) computations\cite{KKR1999}
in 10\% Pb doped Bi2201 to independently verify that the VCA provides
a good description and that Bi/Pb substitution causes little disorder
induced smearing of states.

\begin{figure}
\begin{center}
    \resizebox{8.5cm}{!}{\includegraphics{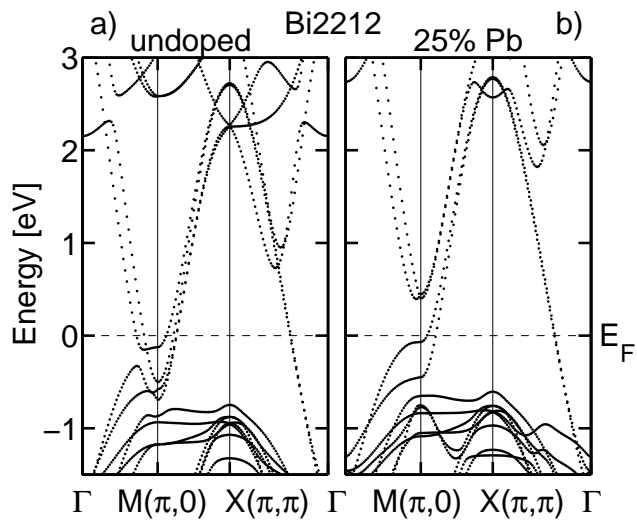}}
\end{center}
\caption{
   \label{fig:1} Band structure of undoped and 25\% Pb-doped
   tetragonal Bi2212 along various high symmetry directions at
   $k_z=0$.}
\end{figure}
We set the stage for our discussion by considering Fig.~1(a) which
shows the band structure of undoped Bi2212 predicted by the LDA. Here
the lattice is assumed to be tetragonal and the structural parameters
used are those of Ref.~\onlinecite{Bellini2004}, obtained by
minimizing the total energy. A pair of closely placed bands is seen to
disperse rapidly through $E_F$ along the $\Gamma-X(\pi,\pi)$ line on
the right side of Fig.~1(a).  These are the well-known CuO$_2$-bands
which are split into bonding and antibonding combinations due to
intracell interactions between the two CuO$_2$-planes. The problem
however is that additional bands of BiO-character drop below $E_F$ at
the $M(\pi,0)$ point giving the so-called `Bi-O pockets', leading to a
metallic Bi-O layer,\cite{Zhang1992} in clear disagreement with
experimental observations\cite{damascelli03}.

Fig.~1(b) shows how the band structure changes dramatically around the
$M$-point when 25\% Pb is substituted for Bi in Bi2212, where the band
structure of the doped compound is computed within the
VCA.\cite{VCAdetail} The Bi-O pocket problem is cured as the Bi-O
bands are lifted to $\approx$ 0.4~eV above $E_F$, and the band
structure around the $M$-point is simplified and the bilayer splitting
of the CuO$_2$ bands around $M$ becomes more clearly visible. The
extended van Hove singularities (VHSs) in the antibonding and bonding
bands appear at binding energies of -0.07~eV and -0.45~eV,
respectively, and the bare bilayer splitting at $M$ is $\approx$
400~meV. The shape of the antibonding and bonding CuO$_2$-bands is
very similar to the generally accepted shape in the cuprates. The
bands in Fig.~1(b) closely resemble the bands obtained in previous
computations\cite{Bansil1999,Lindroos2002} where an {\em ad hoc}
modification of the LDA potential was invoked to account for the
absence of Bi-O pockets in the ARPES spectra of Bi2212.  Further
computations for a range of Pb-doping levels indicate that the Bi-O
pockets are lifted just above $E_F$ at around 22\% Pb-doping in the
tetragonal structure.

\begin{figure}
\begin{center}
    \resizebox{8.5cm}{!}{\includegraphics{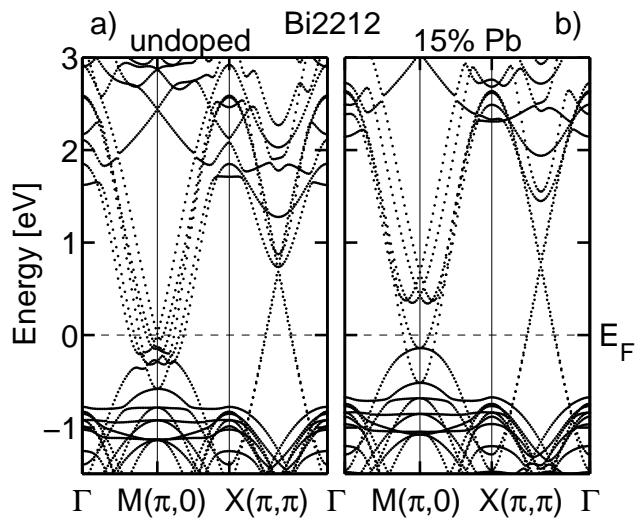}}
\end{center}
\caption{
   \label{fig:2} Band structure (at $k_z=0$) of undoped and 15\%
   Pb-doped Bi2212 assuming orthorhombic lattice structure. Bands are
   plotted along the high symmetry lines of the tetragonal lattice for
   ease of comparison with the results of Fig.~1.}
\end{figure}
The crystal structure of Bi2212 is more realistically modeled as a
$\sqrt{2} \times \sqrt{2}$ orthorhombic unit cell\cite{Sunshine1988}.
Accordingly, Fig.~2 delineates the effect of doping in orthorhombic
Bi2212.\cite{ortholattice,Bellini2004,Miles1998} Here one obtains
twice the number of bands compared to tetragonal Bi2212 due to the
larger size of the unit cell. Comparing Figs.~2(a) and~(b), we see
that 15\% Pb doping in the orthorhombic case lifts the Bi-O pockets
$\approx$ 0.4~eV above $E_F$ and yields a FS consisting of only the
bonding and antibonding CuO$_2$ sheets.

A comparison of Figs.~1 and~2 reveals interesting differences between
the band structures of tetragonal and orthorhombic Bi2212 and their
evolution with Pb doping. The Bi-O complex of bands is more spread out
in energy in Fig.~2(a) than in~1(a), which reflects the larger atomic
displacements in Bi-O layers in the orthorhombic structure. The Bi-O
bands display greater sensitivity to Pb doping in the orthorhombic
case and only 12\% Pb doping pushes the Bi-O pockets above $E_F$
compared to the value of 22\% needed in the tetragonal
structure. There also are differences in the CuO$_2$ bands. For
example, the doping level at which the VHS of the antibonding band
lies at the $E_F$ is 22\% in the orthorhombic structure and 27\% in
the tetragonal case. Besides the highly dispersive CuO$_2$ bands, the
complex of Cu-O bands below $E_F$ (starting around a binding energy of
$\approx$~0.8 eV in Fig.~2(a)), which is primarily composed of Cu $d$
and O $p$ bands, is also influenced by the crystal structure and
doping as seen with reference to Figs.~1 and~2. In particular, in the
doped orthorhombic system in Fig.~2(b), around the $M$-point, these
lower lying bands mix significantly with the CuO$_2$ band involved in
producing the bonding FS sheet, and change the shape of the associated
VHS.

\begin{figure}
\begin{center}
    \resizebox{8.5cm}{!}{\includegraphics{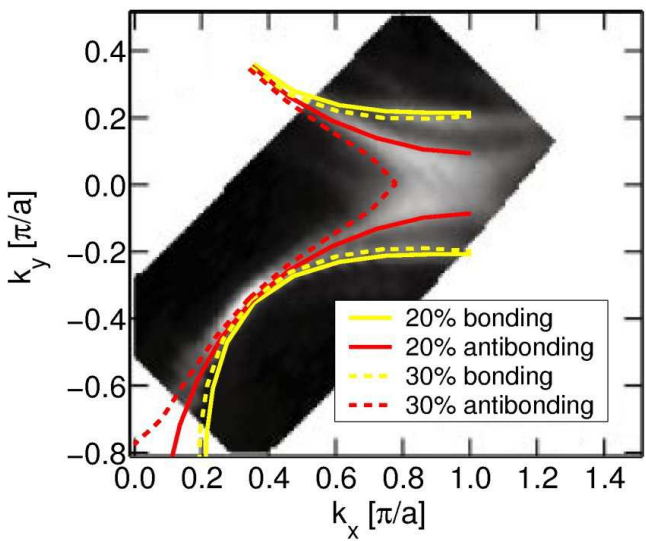}}
\end{center}
\caption{
   \label{fig:3} (Color) Theoretical bonding (yellow lines) and
   antibonding (red lines) FS's of orthorhombic Bi2212 for 20\% (solid
   lines) and 30\% (dashed lines) Pb doping. Other `shadow' FS's are
   not shown for simplicity.  Experimental FS map taken via ARPES from
   an overdoped single crystal of Bi2212 is from
   Ref.~\onlinecite{Bogdanov2001}.  }
\end{figure}
Fig.~3 shows that our theoretically predicted FS is in remarkable
accord with the experimentally determined FS of an overdoped Bi2212
sample obtained via angle-resolved photoemission (ARPES)
measurements\cite{Bogdanov2001}.  For this purpose, computed FS
contours for Pb doping levels of 20\% (solid lines) and 30\% (dashed
lines) for the orthorhombic lattice are overlayed on the experimental
FS map.\cite{kz,Bansil2005,sahrakorpiXX} The `shadow' FS's in the
computations are not shown in order to highlight the main bonding and
antibonding FS's.  The computed bonding FS (yellow lines) shows
relatively little change over 20-30\% doping range and its shape and
dimensions are in quantitative accord with measurements. The
antibonding FS (red lines), on the other hand, is more sensitive to
doping and changes from being hole-like at 20\% doping (solid red
line) to turning electron-like (dashed red line) at 30\% doping as the
$E_F$ descends through the VHS. Therefore the spectral intensity
associated with the antibonding FS in the antinodal region will be
sensitive to local variations in hole doping and a careful modeling of
the spectral intensities will be required to pin down details of the
FS. However, along the nodal direction, neither the antibonding nor
the bonding FS is sensitive to doping and here there is good accord
between theory and experiment.

The driving mechanism underlying the lifting of the Bi-O pockets with
Pb doping in our computations may be understood as follows. The
ionization of Bi atoms in the system will in general generate electric
fields which tend to attract electrons back into the Bi-O layers and
compete with the affinity for the electrons towards the CuO$_2$
layers. The band structures of Figs.~1(a) or~2(a) which display
partially filled Bi-O bands and the associated Bi-O pockets at the FS
then imply that the balance of forces in the computations is such that
Bi is not fully ionized to 3+ in pure Bi2212 so that we may think of
some of the Bi$^{3+}$ electrons as being attracted back to the Bi-O
layers or that the CuO$_2$ layers are self-doped with holes. The fact
that the Bi-O bands are moved above $E_F$ with Pb doping in Figs.~1(b)
or~2(b) then indicates that the substitution of Bi with Pb and the
concomitant reduction of positive charge in the Bi-O layers eliminates
the need for electrons to `flow back' to the Bi-O layer.  In effect
then at e.g. 25\% Pb doping an empty (Bi/Pb)-O band only donates 0.75
rather than 1.0 electron to the CuO$_2$ layer. It is helpful as well
to see how this argument plays out in reverse, i.e. with decreasing Pb
doping. When Pb doping decreases, the tendency of the Bi/Pb electrons
to flow back to the (Bi/Pb)-O layer increases and below a critical
Pb-doping level some of the Bi/Pb electrons actually flow back as the
(Bi/Pb)-O band drops below $E_F$. As a result of this self-doping
effect, further decrease in hole doping of the CuO$_2$ layers is
prevented.

We turn now to consider the effect of adding excess O in Pb-free
Bi2212 for hole doping the system. For this purpose, we have carried
out extensive computations where O, F, or other pseudo-atoms are
inserted in the empty spaces between the Bi-O layers in order to
capture varying amounts of Bi electrons to form a closed
shell\cite{footdelta}. Typical modifications in the band structure are
shown in Fig. 4, where we see that the effect of excess O is to lift
the Bi-O pockets much like that of Pb/Bi substitution. The key is to
reduce the effective number of electrons available in the Bi-O layers
for donation to the CuO$_2$ layers and this can be accomplished via
either Pb/Bi substitution or by adding excess
O.\cite{footsolubility,Majewski2000,MacManus1998} In Fig. 4, the Bi-O
pockets lie below $E_F$ for excess O value of $\delta$= 0.1, but lie
well above $E_F$ for $\delta$= 0.3. Our analysis indicates that the
Bi-O pockets move through the $E_F$ at $\delta\approx$ 0.18.

\begin{figure}
\begin{center}
    \resizebox{8.5cm}{!}{\includegraphics{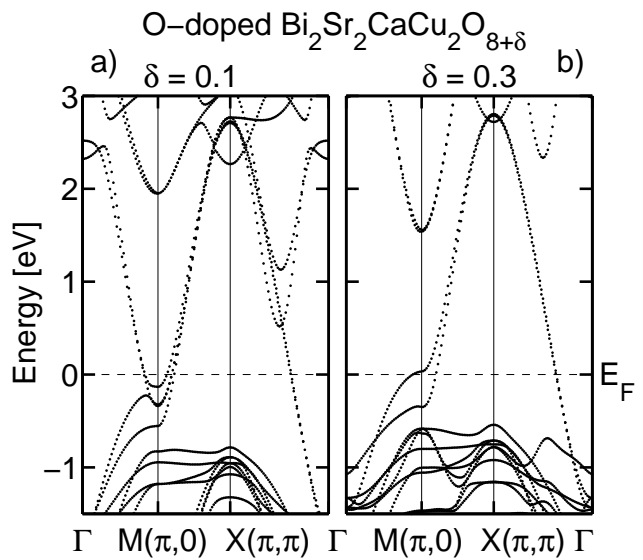}}
\end{center}
\caption{
   \label{fig:4} Band structures of O-doped Bi2212: (a) $\delta =
   0.1$, and (b) $\delta = 0.3$, where $\delta$ denotes excess O per unit
   cell. }
\end{figure}
We have also considered the effect of hole doping on other
Bi-compounds and find that the Bi-O bands in the 15\% Pb doped Bi2201
and Bi2223 are lifted above $E_F$, even though the band structures of
the undoped compounds in both cases display Bi-O FS
pockets.\cite{bi2201lattice,Torardi1988,Miehe1990} Going beyond the
Bi-compounds, we have studied doping effects on the band structures
and binding energies of the core levels \cite{footcore,Tanaka2001} in
the Tl- and Hg-based cuprates. In particular, the Tl-O pockets around
$\Gamma$-point in Tl-based cuprates are removed by approximately 10\%
hole doping.\cite{Plate2005} Interestingly, in Tl$_2$Ba$_2$CuO$_6$
(Tl2201), inclusion of hole doping results in a 3D FS that agrees well
with angular magnetoresistance oscillation measurements of
Ref.~\onlinecite{Hussey2003}. The Hg-derived bands in the Hg-based
cuprates are similarly lifted by O-doping.

In order to assess the effect of superlattice modulation, we have
carried out a series of computations in the orthorhombic structure
where the refined parameters of Miles {\it et al.}  \cite{Miles1998}
at various positions along the modulation were used. We found that the
Bi-O pockets could not be lifted for any value of the
parameters. \cite{distortions,Bellini2004} Even otherwise, we would
not expect superlattice modulation to provide a generic mechanism for
lifting the cation-derived bands in the cuprates because these
modulations vary greatly between different cuprates, and are
suppressed by Pb doping in the Bi-compounds.

In conclusion, our results show clearly that substantial and generic
Coulombic effects come into play with hole doping to lift the
cation-derived bands in the cuprates.  In adducing various physical
quantities from spectroscopic data (e.g. size of the pseudogap),
changes in the electronic structure with underdoping, especially near
the anti-nodal point, should be accounted for, even though most of the
existing analysis in the cuprate literature assumes a
doping-independent band structure. Finally, our study provides a
first-principles route for exploring self-doping effects and doping
dependencies of the electronic structures of this exciting class of
materials.

\begin{acknowledgments}

   We are grateful to S. Kaprzyk for important discussions and
   technical help and thank Z. -X. Shen for bringing
   Ref.~\onlinecite{Tanaka2001} to our attention. This work is
   supported by the USDOE contract DE-AC03-76SF00098, and benefited
   from the allocation of supercomputer time at the NERSC and
   Northeastern University's Advanced Scientific Computation Center
   (ASCC).

\end{acknowledgments}


\begin{thebibliography}{99}
\bibitem{VCAdetail}
%
$Z$ of the effective Bi atom is reduced from 83 to 82.75
to reflect the average charge of the Bi and Pb atoms.
\bibitem{footdelta}
%
An added O atom with Z=8 ($\delta$=1.0) captures two additional
electrons/unit cell from the Bi-O layers to form a closed shell with
Z=10, while an F atom only takes away one electron corresponding to
$\delta$=0.5. A Ne atom (Z=10) possesses a closed shell and we find it
to have little effect on band structure. The values of $\delta$ in
results of Fig.~4 were simulated by adding two pseudo-atoms/unit cell
with Z=10-$\delta$. Very similar results are obtained if O atoms in
Bi-O layers are replaced by pseudo-atoms with Z less than 8.0, so that
these atoms attract more than two electrons.
\end{thebibliography}
\end{document}